\documentclass[aps,prb,floatfix,reprint,twocolumn,reprint,amsmath,amssymb,superscriptaddress,showpacs]{revtex4-1}
\usepackage{graphicx}
\usepackage{bm}
\usepackage{amsmath}
\usepackage{dcolumn}
\usepackage{dsfont}
\usepackage{amssymb}
\usepackage{tabularx}
\usepackage{array}
\usepackage{float}
\usepackage{color}
\usepackage{epstopdf}
\usepackage{mathrsfs}
\usepackage[colorlinks, linkcolor=blue,anchorcolor=blue,citecolor=blue,urlcolor=blue]{hyperref}

\begin{document}

\title{Tunable Hyperbolic Dispersion and Negative Refraction in Natural Electride Materials}

\author{Shan Guan}
\affiliation{Beijing Key Laboratory of Nanophotonics and Ultrafine Optoelectronic Systems, School of Physics,
Beijing Institute of Technology, Beijing 100081, China}
\affiliation{Research Laboratory for Quantum Materials, Singapore University of Technology
and Design, Singapore 487372, Singapore}

\author{Shao Ying Huang}
\email{huangshaoying@sutd.edu.sg}
\affiliation{Engineering Product Development, Singapore University of Technology
and Design, Singapore 487372, Singapore}

\author{Yugui Yao}
\email{ygyao@bit.edu.cn}
\affiliation{Beijing Key Laboratory of Nanophotonics and Ultrafine Optoelectronic Systems,
School of Physics, Beijing Institute of Technology, Beijing 100081, China}

\author{Shengyuan A. Yang}
\email{shengyuan\_yang@sutd.edu.sg}
\affiliation{Research Laboratory for Quantum Materials, Singapore University of Technology
and Design, Singapore 487372, Singapore}

\date{\today}

\begin{abstract}
Hyperbolic (or indefinite) materials have attracted significant attention due to their unique capabilities for engineering electromagnetic space and controlling light propagation. A current challenge is to find a hyperbolic material with wide working frequency window, low energy loss, and easy controllability. Here, we propose that naturally existing electride materials could serve as high-performance hyperbolic medium. Taking the electride Ca$_2$N as a concrete example and using first-principles calculations, we show that the material is hyperbolic over a wide frequency window from short-wavelength to near infrared. More importantly, it is almost lossless in the window. We clarify the physical origin of these remarkable properties, and show its all-angle negative refraction effect. Moreover, we find that the optical properties can be effectively tuned by strain. With moderate strain, the material can even be switched between elliptic and hyperbolic for a particular frequency. Our result points out a new route toward high-performance natural hyperbolic materials, and it offers realistic materials and novel methods to achieve controllable hyperbolic dispersion with great potential for applications.
\end{abstract}


\maketitle
\section{Introduction}
Hyperbolic (or indefinite) medium refers to a class of materials for which the elements of the permittivity or permeability tensor are not of the same sign~\cite{Smith2003}. Such materials are attracting significant interest, because, in contrast to conventional elliptic media, they possess a special hyperbolic-type light dispersion, which leads to many extraordinary optical properties, such as all-angle negative refraction\cite{Smith2004,Hoffman2007,Yao2008}, subwavelength imaging and focusing\cite{Liu2007,Lu2008,Rho2010}, and strong enhancement of spontaneous emission\cite{Jacob2012,Yang2012,Cortes2012}.

The idea of hyperbolic medium was first explored in artificially engineered metamaterials\cite{Poddubny2013}, e.g., with nanowire arrays, metal-dielectric multilayers, and resonant semiconductor heterostructures. This represents a versatile approach, based on which several unique features of hyperbolic medium have been successfully demonstrated. However, the current hyperbolic metamaterials still have several drawbacks\cite{Poddubny2013}: each designed structure normally works well only for a narrow frequency window; the length scale of the structure limits the operating range to long wavelengths, typically from radio frequency to infrared; and more importantly, the energy loss is usually high, both due to strongly absorbing resonances and from scattering off internal structural imperfections.

It was later realized that it could be possible to achieve hyperbolic dispersion also in natural materials with strong structural anisotropy\cite{Sun2014}. Several layered materials, like graphite, MgB$_2$, cuprate, ruthenate, and tetradymite, have been proposed to be hyperbolic\cite{Sun2011,Alekseyev2012,Sun2014,Esslinger2014,Korzeb2015}. Compared with metamaterials, natural hyperbolic materials obviously avoid the need of complicated nanofabrication process and are often available with large samples with less imperfections. Due to the atomic scale periodicity, the operating frequency window can be much wider, and the photon density of states is also significantly larger than metamaterials. With the elimination of dissipation associated with fabrication imperfections, the energy loss is also generally less than metamaterials, nevertheless, it is still quite sizable and remains a severe challenge for any practical application with hyperbolic materials\cite{Sun2014}.

How could we find an ideal hyperbolic material with a wide operating frequency window and with minimal loss? As a general guiding principle, the material must have exceedingly strong anisotropy. Indeed, the indefinite permittivity (or permeability) tensor just reflects an inherent strong anisotropy of the medium. Focusing on indefinite permittivity, a desired scenario would be that free electron motion is confined in one or two spatial directions. In addition, to minimize the energy loss, absorption due to interband optical transitions needs to be suppressed in the working frequency window.

In this work, we propose to achieve high performance hyperbolic medium with a family of natural materials---the electrides. Electrides are a special kind of ionic compound with free electrons serving as the anions\cite{Dye2003}. These electrons are spatially confined and separated from the cations in the crystal, and are not bound to any particular lattice site. Depending on the dimensionality of the anionic confined region, electrides are classified as 0D (confined in isolated cavities), 1D (confined along tubes), and 2D (confined in a planar region)\cite{Dye2009}. One immediately notices that 1D and 2D electrides satisfy the above criterion of strong electronic anisotropy, hence they could be suitable candidates for hyperbolic medium. Here, taking the newly discovered natural electride material Ca$_2$N as a concrete example\cite{Lee2013}, we investigate its optical properties based on first-principles calculations. We show that the extremely strong anisotropy from the 2D confinement of anionic electrons in Ca$_2$N leads to indefinite permittivity in a wide window from short-wavelength to near infrared (0.4 to 1.4 eV). More importantly, the large energy separation between the anionic electron band and the other covalent bonding/anti-bonding bands suppresses the interband optical transitions in the hyperbolic window, resulting in negligible loss due to absorption. The low-loss character is rare and makes this material outstanding among the hyperbolic materials discovered to date. We map out its hyperbolic equifrequency contour, and explicitly demonstrate its negative refraction using numerical simulations. Furthermore, we show that lattice strain offers a convenient method to effectively tune the optical properties of Ca$_2$N. With moderate uniaxial strain, the dispersion can even be switched between elliptic and hyperbolic for a particular frequency. As a result, an incident light with this frequency will undergo a transition from positive to negative refraction. Our result reveals electrides as a highly promising platform to explore high performance hyperbolic medium and the associated many fascinating photonic applications.

\section{COMPUTATIONAL DETAILS}\label{section:methods}
Our first-principles calculations are based on the density functional theory (DFT) using the projector augmented wave (PAW) method\cite{Bloechl1994} as implemented in the Vienna $ab$-$initio$ simulation package (VASP)\cite{Kresse1993,Kresse1996}. For structural optimization, The generalized gradient approximation with the Perdew-Burke-Ernzerhof (PBE)\cite{Perdew1996} realization is adopted for the exchange-correlation functional. The cutoff for plane wave expansion is set to be 600 eV. The lattice constants and the ionic positions are fully optimized until the force on each ion is less than 0.01 eV/\AA. The convergence criteria for energy is chosen as 10$^{-5}$ eV. DFT-D2 method is applied to describe the long-range van der Waals interaction\cite{Grimme2006}. Monkhorst-Pack $k$-point mesh with size of 25$\times$25$\times$5 is used for geometry optimization. To achieve highly converged optical results, the size of $k$-mesh is increased to 35$\times$35$\times$7 for calculating permittivity. The more accurate hybrid functional (HSE06)\cite{Heyd2003} approach is adopted for calculating the electronic band structure and optical properties. In calculating the permittivity, the experimentally measured electron lifetime of 0.6 ps is used for evaluating the intraband Drude contribution\cite{Lee2013}. On applying the uniaxial strain, all atomic positions and the lattice constant normal to the strain direction are fully relaxed according to the stated energy and force convergence criteria.

\section{RESULTS AND DISCUSSION}\label{section:results}
\begin{figure*}[htb!]
\centerline{\includegraphics[width=0.8\textwidth]{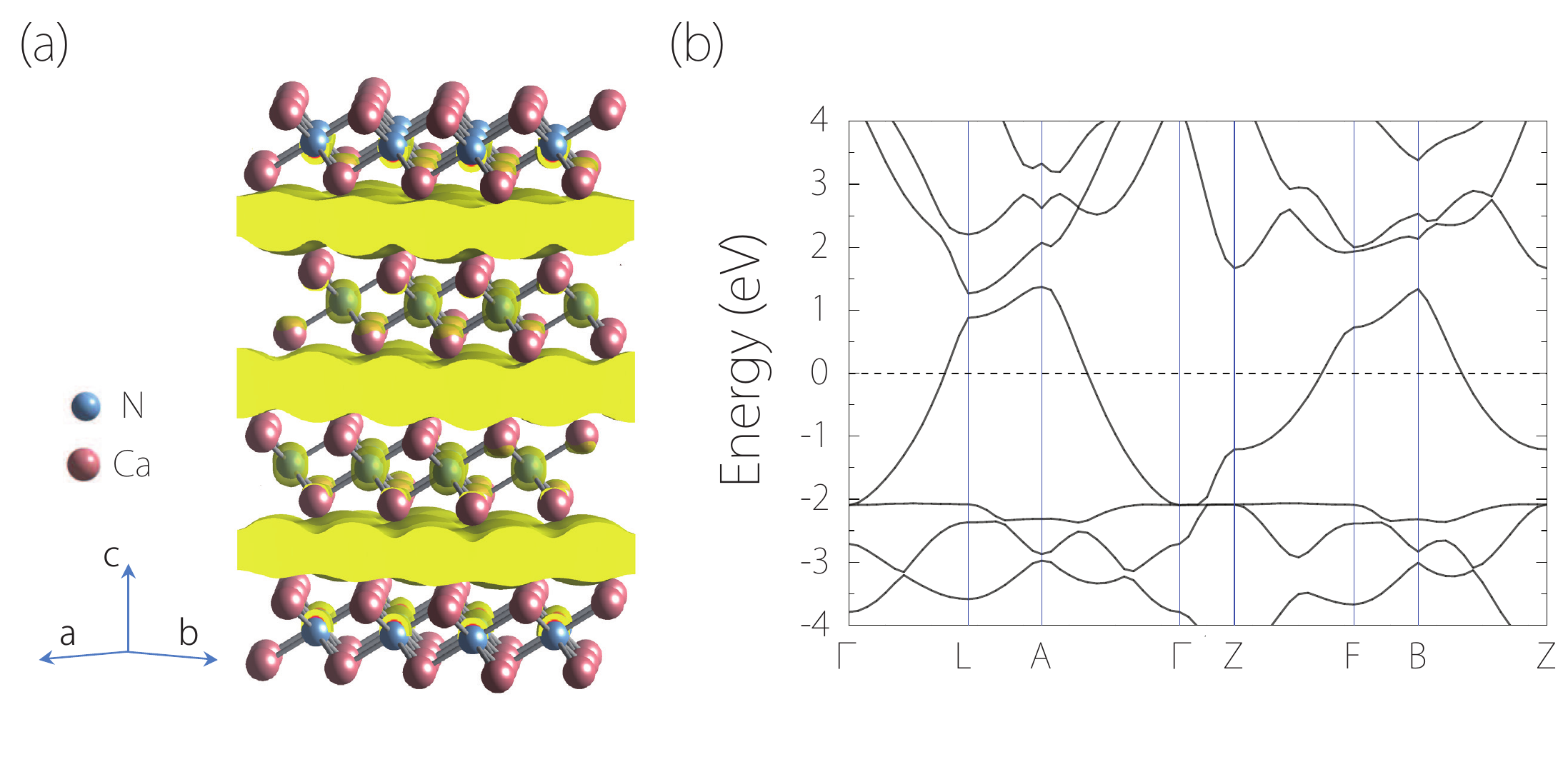}}
\caption{(a) Lattice structure of Ca$_2$N crystal, along with the electron density distribution for states around Fermi energy (within an energy window of 0.1 eV). (b) Electronic band structure of Ca$_2$N. The band that crosses the Fermi level is from the anionic electrons mostly distributed in the interlayer regions (as illustrated in (a)). }\label{fig1}
\end{figure*}

Ca$_2$N crystal has a rhombohedral layered structure with $R\bar{3}m$ space group (anti-CdCl$_2$-type), in which the three covalently bonded atomic layers are grouped into a Ca-N-Ca triple layer (TL), and the TLs are weakly bonded and are stacked along the crystal $c$-axis (denoted as the $z$-direction here) to form a three-dimensional crystal as shown in Fig. \ref{fig1}{\color{blue}(a)}. Using first-principles calculations based on the density functional theory (DFT) (see Methods section for details), we obtain the optimized equilibrium lattice constants $a=3.55$ \AA\, and $c=18.85$ \AA\, (for the conventional hexagonal unit cell), which agree well with the experimental values ($a=3.62$ \AA\, and $c=19.10$ \AA)\cite{Reckeweg2002}and other theoretical results\cite{Inoshita2014}.

Counting the valence of the elements, one observes that there is one excess electron per formula unit. This electron gets largely confined in the planar regions between the TLs, forming the anionic electron layers and making Ca$_2$N a 2D electride material\cite{Lee2013,Oh2016}. This message is usually conveyed by writing the chemical formula as [Ca$_2$N]$^+\cdot$e$^-$. It has been argued that the 2D electride character underlies the unltra-low work function of Ca$_2$N\cite{Uijttewaal2004}. It was further predicted that single- or few-layer Ca$_2$N could be exfoliated and may behave as good electronic or battery electrode materials\cite{Zhao2014,Guan2015,Hu2015a}. Here our focus is on bulk Ca$_2$N. It should be re-emphasized that the adjective "2D" in "2D electride" refers to the topology of the anionic confinement region, it does not mean the material itself is two-dimensional (rather, it is a 3D material here).

To illustrate this unique feature, we calculate the electronic band structure of Ca$_2$N and the result is plotted in Fig. \ref{fig1}{\color{blue}(b)}. Here in order to achieve better accuracy for subsequent study on the optical properties, we adopt the hybrid functional (HSE06)\cite{Heyd2003} approach for calculating the band structure. From Fig. \ref{fig1}{\color{blue}(b)}, one observes that a band crosses the Fermi level with a large bandwidth $>3$ eV. This band is mainly contributed by the anionic electrons between the TLs, as can be clearly seen in Fig. \ref{fig1}{\color{blue}(a)} of the charge density distribution for states around the Fermi level. The band is highly dispersive in the $xy$-plane and less dispersive along the $z$-direction, indicating that these anionic electrons are much more freely to move in the interlayer region than to move across the TLs. Such exceedingly strong electronic anisotropy underlies the appearance of indefinite permittivity over a wide range of spectrum. Moreover, another important observation is that the states around the Fermi level are solely from the anionic electron band, whereas other bands from the covalently bonded electrons are pushed far away from the Fermi level ($<-2$ eV or $>1.5$ eV). This feature plays a crucial role in the low energy loss to be discussed in the following.

\begin{figure}[htb!]
\centerline{\includegraphics[width=0.4\textwidth]{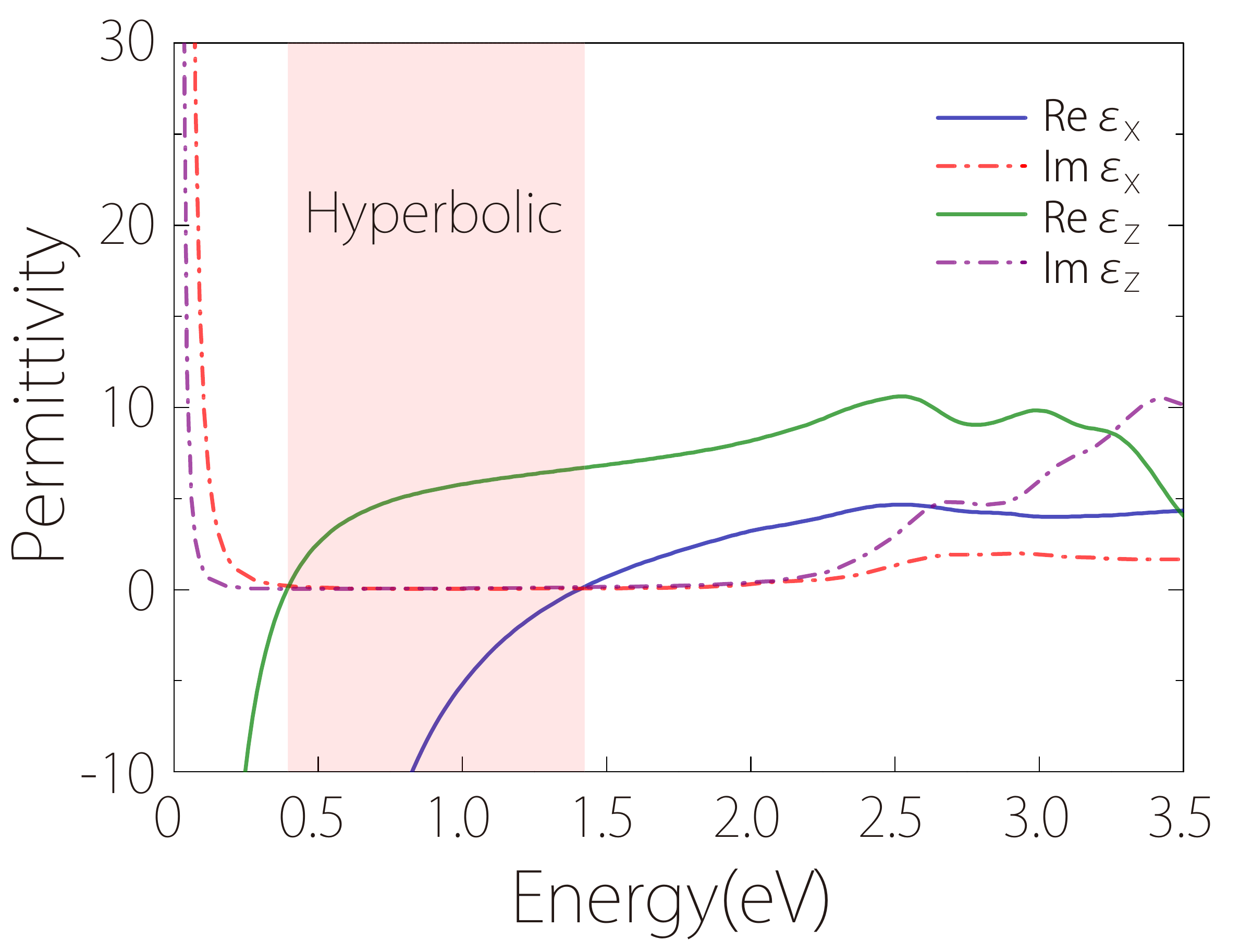}}
\caption{Real and imaginary parts of the two principal components of permittivity of Ca$_2$N. The shaded region shows the window in which the material is hyperbolic.}\label{fig2}
\end{figure}

Based on the band structure, both the imaginary and the real part of the permittivity can be evaluated via the standard approach\cite{Harl2007}. Due to the $D_{3d}$ point group symmetry, the permittivity tensor has only two independent elements $\varepsilon_{x}(\omega)$ and $\varepsilon_{z}(\omega)$ along the principal axis. Their real and imaginary parts are plotted in Fig. \ref{fig2}. First of all, one observes that the real part of each element is negative at low frequency and crosses zero from negative to positive values with increasing frequency, as a characteristic of metallic material. Importantly, due to the strong anisotropy, they cross zero at frequencies with a large difference: about $0.38$ eV for $\varepsilon_{z}$ and $1.40$ eV for $\varepsilon_{x}$. This is consistent with the observation that the anionic electrons are more freely to move in the $xy$-plane. As a result, $\varepsilon_{x}$ and $\varepsilon_{z}$ have different signs between $0.38$ eV and $1.40$ eV, and the material becomes hyperbolic in this wide frequency window. Meanwhile, the imaginary parts of $\varepsilon_{x}$ and $\varepsilon_{z}$ are associated with transitions and directly determine the energy loss due to absorption. They exhibit characteristic peaks around zero frequency due to intraband transitions close to the Fermi level (known as Drude peaks). Remarkably, above the Drude peak, both imaginary parts get almost completely suppressed over a large frequency range from about 0.4 eV up to 2 eV, covering the hyperbolic window. This means that there is almost no energy loss when the material is hyperbolic. The behavior can be understood by invoking the previous observation of the large energy separation between the single anionic electron band and other covalent bonding bands. Since optical transitions are vertical (i.e., at the same wave vector) and are between states of different occupation (i.e., across the Fermi level), the large band separation diminishes the interband optical transitions for excitation energies below about 2 eV. Thus, the result indicates that Ca$_2$N is hyperbolic and nearly lossless over a wide frequency window from short-wavelength to near infrared (from about 880 nm to 3 $\mu$m).

\begin{figure*}[!htb]
\begin{center}
\centerline{\includegraphics[width=0.8\textwidth]{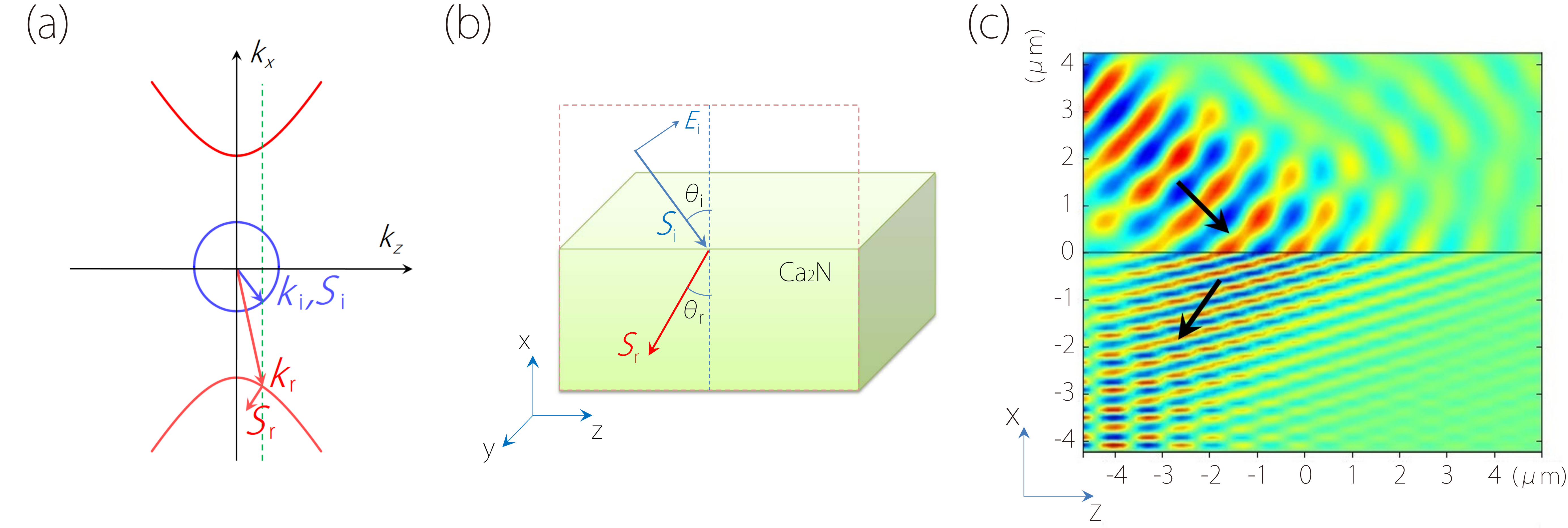}}
\caption{ (a) Plot of equifrequency contours (EFCs) followed from Eq .(\ref{EFC}). The blue circle is the EFC of air, and the red hyperbola is the EFC of Ca$_2$N. Here we take a light frequency of 1.179 eV. (b) Negative refraction occurs when a TM polarized light is incident from air towards Ca$_2$N. The blue (red) arrow indicates the direction of Poynting vector $\bm S_i$ ($\bm S_r$) of the incident (refracted) light. Note that the crystal $c$-axis is along the $z$-direction, i.e., parallel to the interface.
(c) Simulation result explicitly showing the negative refraction effect. The setup follows from (b): air (Ca$_2$N) occupies the region of $x>0$ ($x<0$). The light frequency is taken to be 1.179 eV. The incident angle is 45$^\circ$. The colormap shows the electric field strength, and the black arrows indicate the directions of local Poynting vectors.}\label{fig3}
\end{center}
\end{figure*}

The indefiniteness of permittivity has highly nontrivial effect on the light propagation. Consider a transverse magnetic (TM) polarized (\emph{p}-polarized) light incident from air onto the hyperbolic material. Let the crystalline $c$-axis ($z$-direction) be parallel to the interface and the incident plane be the $xz$-plane, as shown schematically in Fig. \ref{fig3}{\color{blue}(b)}. Then the light dispersion within a particular region is given by
\begin{equation}\label{EFC}
\frac{k_x^2}{\varepsilon_z}+\frac{k_z^2}{\varepsilon_x}=\left(\frac{\omega}{c}\right)^2.
\end{equation}
For each fixed frequency $\omega$, this relation gives an equifrequency contour (EFC) in the wave vector plane. As shown in Fig. \ref{fig3}{\color{blue}(a)}, the EFC is a circle in air due to the isotropic permittivity. (For definite medium with anisotropy, it is an ellipse.) In contrast, in the hyperbolic material with $\varepsilon_x$ and $\varepsilon_z$ being of different signs, the EFC becomes a hyperbola (which is the reason for the term "hyperbolic"). For the case of Ca$_2$N, we have $\varepsilon_z>0$ and $\varepsilon_x<0$ in the hyperbolic window, hence the EFC is a hyperbola with focal points on the $k_x$-axis and has an eccentricity of $\sqrt{1-\varepsilon_x/\varepsilon_z}$ (see Fig. \ref{fig3}{\color{blue}(a)}).

Many unusual properties of hyperbolic material directly follows from this hyperbolic EFC. For example, the propagating modes in conventional elliptic (definite) materials have a finite range of allowed wave vectors, beyond which the waves are evanescent and contain sub-wavelength information. This can be easily seen from Fig. \ref{fig3}{\color{blue}(a)} where the EFC in air (or any other definite material) is bounded with a finite $k_z$. In sharp contrast, a hyperbolic EFC has no bound in $k_z$, which suggests that propagating modes exist for arbitrarily large wave vectors (eventually cut off by the lattice length scale where the effective medium description breaks down). This allows hyperbolic materials to convert evanescent modes to propagating modes for sub-wavelength imaging, leading to the proposition of hyperlens\cite{Jacob2006}.

Another direct manifestation of hyperbolic EFC is the all-angle negative refraction effect. For the considered configuration in Fig. \ref{fig3}{\color{blue}(b)}, the wave vector component parallel to the interface (i.e., $k_z$) is conserved during refraction. For a given incident wave vector $\bm k_i$, this condition determines the wave vector $\bm k_r$ for the refracted wave (see Fig. \ref{fig3}{\color{blue}(a)}). Both wave vectors are on the same side of the interface normal ($x$-direction). The group velocity of the light, which is in the direction of the Poynting vector, points to the direction normal to the EFC. In air, the Poynting vector $\bm S_i$ is along $\bm k_i$. However, in the hyperbolic material, $\bm S_r$ is always on the opposite side of the interface normal (with respect to $\bm k_r$ and also $\bm S_i$)---a direct consequence of the hyperbolic EFC (see Fig. \ref{fig3}{\color{blue}(a)}), leading to negative refraction for all incident angles. For a given angle of incidence $\theta_i$, the refraction angle can be easily found as
\begin{equation}\label{angle}
\theta_r=\arctan\left[\frac{\sqrt{\varepsilon_z}\sin\theta_i}{\varepsilon_x\sqrt{1-\frac{\sin^2\theta_i}{\varepsilon_x}}}\right],
\end{equation}
which confirms that $\theta_r$ and $\theta_i$ are of opposite signs for the case with $\varepsilon_z>0$ and $\varepsilon_x<0$.

We explicitly demonstrate the negative refraction effect for Ca$_2$N by performing numerical simulations of the Maxwell equations with finite element method using our calculated permittivity tensor. We take the incident angle to be 45$^\circ$ and the light frequency to be of 1.179 eV, and standard absorbing boundary condition is applied. The configuration follows the schematic setup in Fig. \ref{fig3}{\color{blue}(b)}. The air/Ca$_2$N interface is at $x=0$, with $x>0$ ($x<0$) region occupied by air (Ca$_2$N). Fig. \ref{fig3}{\color{blue}(c)} shows the distribution of the electric field, which clearly demonstrate the negative refraction at the interface. Here the black arrows mark the directions of the calculated local Poynting vectors. The obtained refraction angle agrees with Eq .(\ref{angle}).

\begin{figure}[!htb]
\centerline{\includegraphics[width=0.4\textwidth]{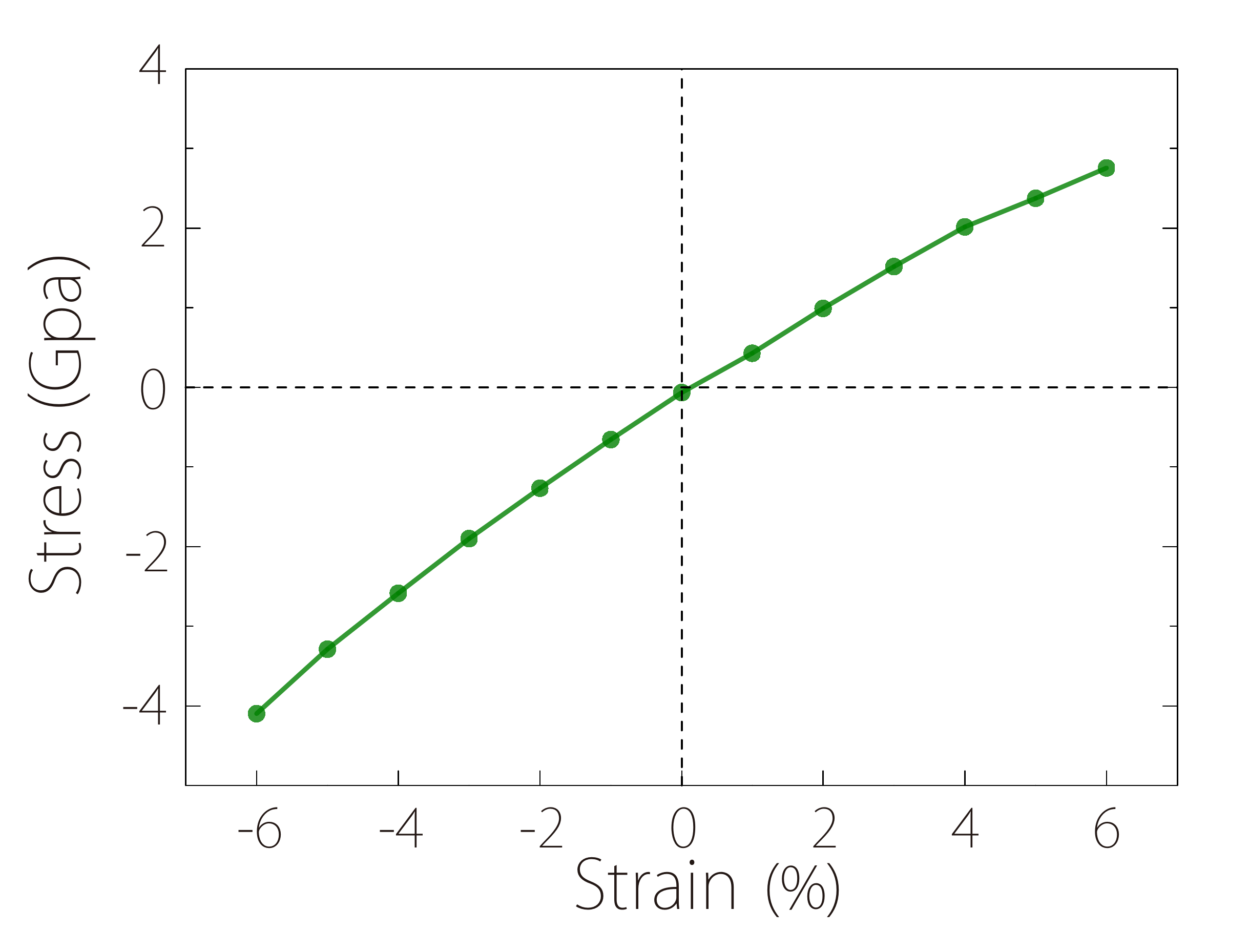}}
\caption{ Strain-stress relation for Ca$_2$N with uniaxial strains along $z$-direction. }\label{fig4}
\end{figure}

\begin{figure*}[!htb]
\centerline{\includegraphics[width=0.8\textwidth]{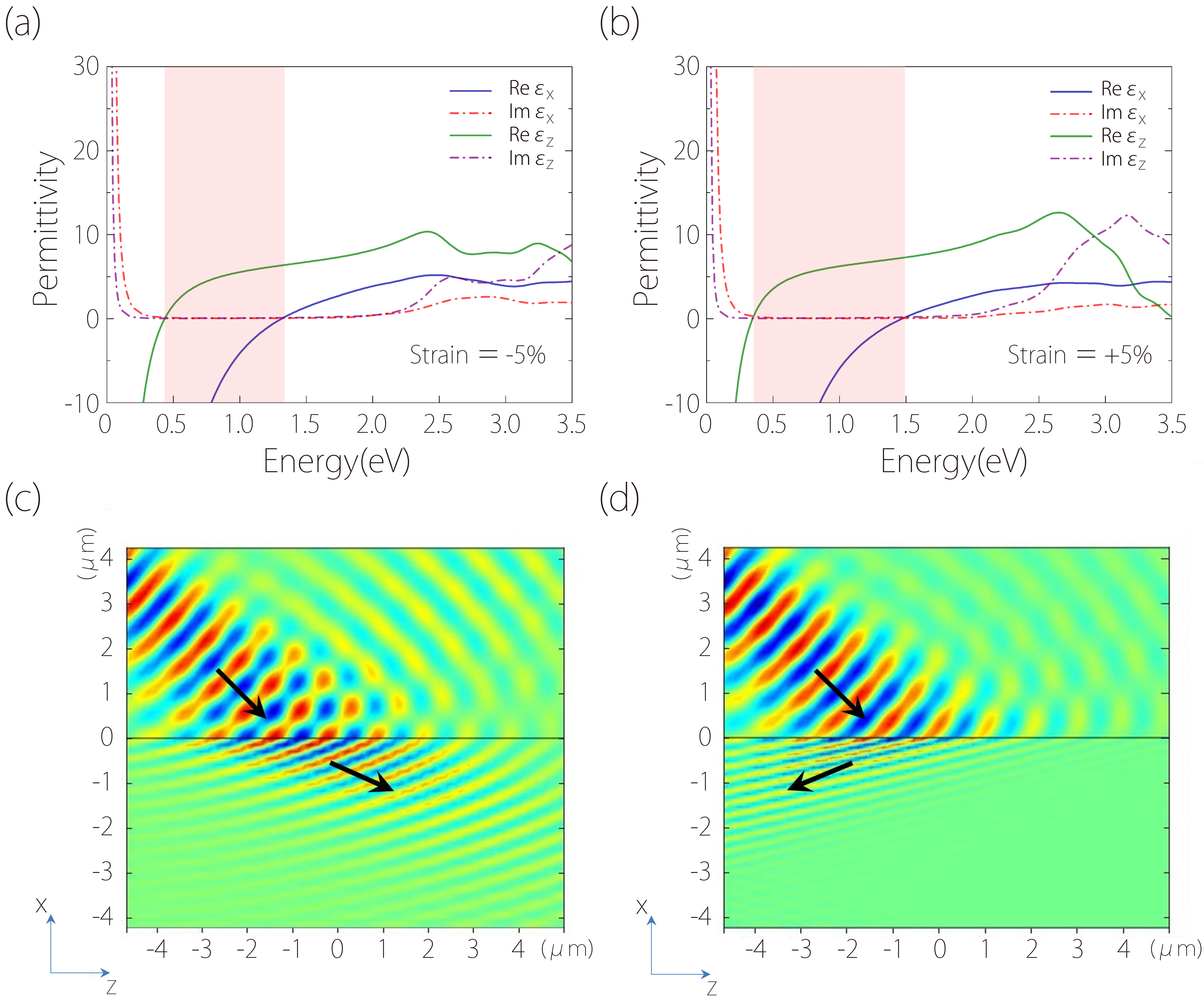}}
\caption{Real and imaginary parts of the permittivity of Ca$_2$N under (a) $-5$\% and (b) $+ 5$\% strains. Note the sizable change in the range of hyperbolic window (the shaded region). (c,d) Simulation results for the refraction process near the air/Ca$_2$N interface with Ca$_2$N under two different strains: (c) $-5$\% and (d) $+ 5$\%. The setup is similar to Fig. \ref{fig3}{\color{blue}(c)}. The light frequency is taken to be 1.453 eV, and the incident angle is 45$^\circ$. The directions of local Poynting vectors are indicated by black arrows.}\label{fig5}
\end{figure*}

The properties of solid state materials can be conveniently tuned by strain. Since Ca$_2$N has a layered structure with large interlayer separation ($>3.5$ \AA), uniaxial strain along the $z$-direction is expected to be an effective approach to control its optical property. We first analyze Ca$_2$N's elastic properties. The calculated strain-stress relation is shown in Fig. \ref{fig4}. One observes that linear elastic regime extends over a wide range, beyond $\pm 6\%$ strain. The Young's modulus is estimated to be about 50 GPa, which is quite small (e.g., compared with 130-185 GPa for Si single crystal)\cite{BoydFeb.2012}, which should facilitate the strain engineering of its property.

We find that the permittivity tensor can be continuously tuned by the applied uniaxial strain. In Fig. \ref{fig5}{\color{blue}(a)} and \ref{fig5}{\color{blue}(b)}, we plot the permittivity for strains $-5\%$ and $+5\%$ respectively. The qualitative features of the curves are the same as the unstrained case. The most important difference is that the frequency window for hyperbolic dispersion is changed. For $-5\%$ strain, this window spans from 0.44 to 1.32 eV; whereas for $+5\%$ strain, the window is further extended from 0.34 to 1.48 eV. Consequently, for frequencies covered in one window but not the other, the corresponding dispersions will change between elliptic and hyperbolic upon the application of strain. For example, consider the same setup as in Fig. \ref{fig3}{\color{blue}(c)} with an incident light with a frequency of 1.453 eV. Under $-5\%$ strain, the dispersion in Ca$_2$N is elliptic for this frequency and the light undergoes normal positive refraction (see Fig. \ref{fig5}{\color{blue}(c)}). Now if we apply a $+5\%$ strain, the dispersion becomes hyperbolic, hence the refraction switches to negative (see Fig. \ref{fig5}{\color{blue}(d)}). Such an interesting effect is indeed confirmed by our numerical simulation results in Fig. \ref{fig5}{\color{blue}(c)} and \ref{fig5}{\color{blue}(d)}. To our best knowledge, this is the first time that the physical approach of strain has been proposed to switch an optical medium between elliptic and hyperbolic. It offers additional flexibility in controlling the light propagation and will open great possibilities for devices that combine mechanical and optical functionalities.

We discuss a few points before concluding. First, as natural materials, the electrides have the advantage over metamaterials that they do not need complicated nano-fabrication process which usually introduce imperfections and strongly absorbing resonances. In particular, high-quality single crystalline Ca$_2$N samples with millimeter scale have been demonstrated\cite{Lee2013}. Due to its layered structure, Ca$_2$N slabs or thin films may also be conveniently obtained via mechanical exfoliation\cite{Lee2013,Druffel2016}. These factors should facilitate the experimental investigation of its optical properties and future applications.

Second, besides the above-mentioned negative refraction and sub-wavelength imaging, the enhanced photonic density of states due to hyperbolic dispersion will also be interesting. For example, it can lead to a large Purcell factor for spontaneous emission. Theoretically, the maximum Purcell factor is on the order of $(\lambda/a)^3$, where $\lambda$ is the light wavelength and $a$ is the lattice parameter of the structure\cite{Poddubny2013}. For natural materials like electrides, $a$ is of atomic length scale, hence the Purcell factor enhancement could be much larger than that of metamaterials. The effect is useful for engineering spontaneous emission and also increasing the efficiency of heat transfer.

Finally, we stress that the hyperbolic behavior should be general for 1D and 2D electride materials, not limited to Ca$_2$N. The presence of 1D or 2D confined anionic electrons represents a strong structural anisotropy, naturally leading to wide hyperbolic frequency window. For example, we find that two other 2D electride materials Sr$_2$N and Ba$_2$N also exhibit hyperbolic behavior. These two electrides share similar layered structures with Ca$_2$N\cite{Walsh2013}. In the Supporting Information, we show that they are also low-loss hyperbolic materials in the infrared frequency range. Sr$_2$N has hyperbolic window from 0.40 to 1.25 eV, and Ba$_2$N has hyperbolic window from 0.31 to 0.96 eV. With the exploration of new electride materials\cite{Inoshita2014,Ming2016}, we expect that the hyperbolic window could be further extended to cover a wide spectrum.

\section{CONCLUSION}\label{section:conclusions}
In summary, we reveal electrides as a promising new family of hyperbolic materials.
Using first-principles calculations and taking the naturally existing Ca$_2$N as a concrete example, we show that
indefinite permittivity can appear in a wide range spanning the short-wavelength to near infrared. This indefinite permittivity is attributed to the extremely strong anisotropy from the 2D confined anionic electrons. Importantly, the energy loss due to absorption is suppressed in this hyperbolic window, making it an ideal candidate for practical applications. In addition, we propose strain engineering as an effective method to control its optical properties. We show that with moderate strain, the material can be switched between elliptic and hyperbolic for a particular frequency. The negative refraction effect and its tunability by strain are explicitly demonstrated with numerical simulations. Considering the important advantages such as the natural availability without any nano-fabrication, the wide hyperbolic frequency window, the rather low energy loss, and the easy control via strain, the Ca$_2$N electride material is expected to be a highly promising platform to explore various fascinating effects of hyperbolic medium and the many possible photonic device applications.

\begin{acknowledgments}
We thank  D. L. Deng and Hongyi Xu for valuable discussions. This work was supported by the MOST Project of China (Nos.2014CB920903, 2016YFA0300603), the National Natural Science Foundation of China (Grant Nos.11574029, and 11225418), and Singapore MOE Academic Research Fund Tier 1 (SUTD-T1-2015004).

\end{acknowledgments}

%

\end{document}